\documentclass[prl,twocolumn,showpacs,showkeys,preprintnumbers,amsmath,amssymb]{revtex4}% Physical Review Letter

 \usepackage{epsf}

\usepackage{amsmath,amsfonts}

\newcommand{\bra}{{\langle}}
\newcommand{\ket}{{\rangle}}

\begin{document}
\preprint{hep-th/0502172, NSF-KITP-05-11}
\title{Quantizing Open Spin Chains with Variable Length: an example from Giant Gravitons}
\author{David Berenstein$^{1,2}$}
\email{dberens@physics.ucsb.edu}
\author{Diego H. Correa$^2$}
\email{dcorrea@kitp.ucsb.edu}
\author{Samuel E. V\'azquez$^1$}
\email{svazquez@physics.ucsb.edu}

 \affiliation{$^1$Department of
Physics, University of California
at Santa Barbara, CA 93106 \\
$^2$Kavli Institute for Theoretical Physics, University of
California at Santa Barbara, CA 93106}

\begin{abstract}
We study an XXX open spin chain with variable number of sites,
where the variability is introduced only at the boundaries. This
model arises naturally in the study of Giant Gravitons in the
AdS/CFT correspondence. We show how to quantize the spin chain by
mapping its states to a bosonic lattice of finite length with
sources and sinks of particles at the boundaries. Using coherent
states, we show how the Hamiltonian for the bosonic lattice gives
the correct description of semiclassical open strings ending on
Giant Gravitons.
\end{abstract}
 \maketitle

{\it I.  Introduction. }

In the last few years many connections have been made relating
large $N$ quantum field  theories in four dimensions, string
theory on negatively curved space-time and quantum spin chain
models. These seemingly disparate subjects have been tied together
via the AdS/CFT correspondence \cite{Malda, GKP98, Witten98, BMN,
minahan}. In particular, it has become apparent that to understand
the field theory beyond the one loop approximation, one needs to
deal with the problem of studying spin chain models with varying
numbers of sites \cite{Beisert}.

In this article we want to report a new development relating to
this collection of subjects for the case of open spin chain
models. In particular we will show how one can quantize an XXX
spin chain model with variable numbers of sites, where the
variability is introduced only at the boundaries. This model makes
its natural appearance in the study of  string states attached to
giant gravitons when seen from the dual CFT point of view, and the
variability on the number of sites is obtained already at one loop
order. The XXX model has also important applications in condensed matter
and statistical physics and here we provide a generalization of the boundary conditions
for a finite length chain.
For the present paper we will content ourselves with the
form of this final answer and present a derivation of the model  and a more complete study elsewhere \cite{BCV}.

We remind the reader the basic geometric properties of
giant gravitons \cite{mcgreeve, grisaru, hashimoto, vijay, corley,
vijay2, david, vijay3}. These are D3-branes of $AdS_5\times S^5$
that wrap an $S^3$ inside the $S^5$, and move in a circle. More
specifically, the radius of their orbit in the $S^5$ is determined
by their angular momentum $p$ as, $r = R \sqrt{1 - p/N}$. Here $R$
is the radius of the $S^5$ and $N$ is the number of units of
five-form flux on the sphere. Hence, the angular momentum is
bounded by $p \leq N$, with the equality corresponding to the
``maximal" giant graviton. Also, the Hamiltonian we discuss here is associated
to the one loop anomalous dimension  of certain operators in the
dual field theory \cite{vijay2,david,vijay3}.

Our main result is that the Hamiltonian of the spin chain model
for the variable number of sites that we consider can be
transformed to a Hamiltonian for a dual model with a fixed number of sites and
non-diagonal boundary conditions. This lets us solve for the
ground state of the model and apply the results to the study of
the Dirac-Born-Infeld fluctuations of the brane. From the string
theory point of view, this Hamiltonian can be understood by a
different gauge choice of the Polyakov action than the standard
one, and it is a hint that the CFT knows something about the
reparametrization invariance of the string.

{\it II. A map from the XXX spin chain model to a system of bosons on a lattice}

Let us begin with a description of the configuration space of a
finite chain of length $L$ of the XXX model. This is, consider a
spin system with $L$ sites, some of  which are set to spin up, and
some which are set at spin down. We will label the spin down site
with a  $Y$, and the spin up site by $Z$. A configuration of spins
of the system can be mapped into a word on the letters $Y,Z$. For
example, we can consider the states  as words $Y^L$, or $Y^k Z
Y^{L-k-1}$. The first one is the ferromagnetic ground state of the
XXX spin chain model, and the second one is a state with one
impurity. We can also build multi-impurity states by inserting $Z$
at various locations.

Now let us assume that neither the first site nor  the last site
is allowed to be a $Z$. In effect, this boundary condition arises
from the giant gravitons. We can consider all
words in $Y,Z$ as being generically of the following form:
\begin{equation}
YZ^{n_1}Y Z^{n_2} Y \dots Y Z^{n_k} Y\;,
\end{equation}
where $L= k+1 +\sum_i{n_i}$ and the $n_i$ are  non-negative
integers (notice there are $k+1$ different $Y$'s in the
expression). The spin $s_z$ of the system relative to the
ferromagnetic ground state is $s_z- s^0_z = \sum_i n_i$. We can
map this state to a state of a lattice with $k$ sites where at
each site $i$ there is a boson with occupation number $n_i$. The
total occupation number is then $\sum_i n_i$. This duality map
does not preserve the length of the spin chain. If we consider the
set of all states of a fixed length $L$ for the XXX model, this
will be mapped to a collection of states of different length $k$
on our bosonic lattice depending on the total spin of the
configuration. Similarly, the set of all boson configurations of
length $k$ gets mapped to a collection of arbitrarily large spin
chain configurations of the XXX model, depending on the total
occupation number.

Now let us consider the standard ferromagnetic XXX spin chain
Hamiltonian. This Hamiltonian preserves the number of $Z$. This
means that the Hamiltonian acting on the states of the bosonic
chain does not mix lengths of the spin chain, and can also be
understood as a lattice model which preserves the boson number.
This turns out to be a nearest neighbor Hamiltonian as well.

To derive the Hamiltonian for the boson chain, we first  define the
 oscillator-like operators, $\hat{a}_i$, $\hat{a}_i^\dagger$ acting
at each site with the property $\hat{a}_i|n_i\ket = |n_i - 1\ket$,
$\hat{a}_i|0\ket = 0$ and $\hat{a}_i^\dagger |n_i\ket |n_i+1\ket$ (These are shift operators for an infinite basis). It
follows that these operators describe a free Fock space for a
single species (e.g. \cite{minic1} and references therein). They
obey the so-called ``Cuntz" algebra for a single species
\cite{Cuntz},
\begin{eqnarray}
\label{ops}
\hat{a}_i \hat{a}_i^\dagger = I\;, \quad
\hat{a}_i^\dagger \hat{a}_i = I - |0\ket\bra 0|\;.
\end{eqnarray}
The number operator $\hat{n}_i$ can also be built from these
operators as in \cite{minic1}. Note that operators at different
sites are commuting,  so ``particles" filling the sites respect
bosonic statistics. This algebra can also be considered as the $q
\rightarrow 0$ limit of the deformed harmonic oscillator algebra
$\hat{a} \hat{a}^\dagger - q \hat{a}^\dagger \hat{a} = 1$.

The Hamiltonian takes a
particularly simple form,
\begin{eqnarray}
\label{H0}
 H = 2
\lambda \sum_{l = 1}^L \hat{a}_l^\dagger \hat{a}_l - \lambda
\sum_{l=1}^{L-1}(\hat{a}_l^\dagger\hat{a}_{l+1} + \hat{a}_l
\hat{a}_{l+1}^\dagger)\;,
\end{eqnarray}
where we have chosen the ferromagnetic ground state to have zero
energy, and for the giant gravitons $\lambda = \frac{g_s N}{2\pi}$
with $g_s$ the string coupling \cite{BMN, minahan}.
 The first term tells us that there
is a finite amount of energy in each oscillator,  which is
$2\lambda$ if the bosonic oscillator is occupied and zero
otherwise (in the XXX model this is twice the energy for a domain
wall between a region of $Z$ followed by one of $Y$). The second
term is interpreted as a hopping term for bosons to move
between sites, so that the energy is reduced with bosons that are
not localized. Clearly, if we consider the collection of all
possible spin chain lengths for both the XXX model and our ``Cuntz
oscillator" model, we obtain a complete identification of the two
dynamical systems.
In the above, we have assumed fairly standard
boundary conditions for the spin chain (it preserves the total
spin), although for the $Z$ in the XXX model we have some sort of
Dirichlet boundary condition, as we are forbidding the $Z$ to be
at the edge of the word. This boundary condition is integrable and
the model is solvable by Bethe Ansatz techniques \cite{sam}.
Similar limits in $q$-boson hopping models have been studied for the case of periodic boundary conditions \cite{BBP}.

{\it III. Non-diagonal boundary conditions}

Let us now consider a new version of the Hamiltonian in the Cuntz oscillator which
corresponds to a non-diagonal boundary condition (here, we refer to the fact that the
total boson occupation number does not commute with the Hamiltonian)
\begin{eqnarray}
\label{H}
 H &=& 2 \lambda \, \alpha^2 + 2
\lambda \sum_{l=  1}^L \hat{a}_l^\dagger \hat{a}_l - \lambda
\sum_{l=1}^{L-1} (\hat{a}_l^\dagger\hat{a}_{l+1} +
\hat{a}_l \hat{a}_{l+1}^\dagger) \nonumber \\
&& + \lambda \,\alpha \,(\hat{a}_1^\dagger + \hat{a}_1) +
\lambda\, \alpha\,(\hat{a}_L^\dagger + \hat{a}_L)\;,
\end{eqnarray}
where $\alpha = \sqrt{1 - p/N}$ with $p$ and $N$  integers (they
are arbitrarily large with $p\leq N$, so their fraction $p/N$ is a
general real number between $0$ and $1$). Notice that when $p/N\to
1$ the result reduces to the Hamiltonian in the previous section.
Turning $p/N$ away from one produces a new boundary condition for
the spin chain. From the point of view of the Cuntz oscillator
algebra, we have a source/sink of bosons at each end. If we go
back to the XXX spin chain, here we have a generalization that
adds and subtracts sites to the model, and corresponds to a spin
chain with a dynamical number of sites. This is the mathematical
problem that we discussed in the introduction. The Hamiltonian we
have written above is the one that can be derived from attaching
strings to a non-maximal giant graviton from the dual CFT
\cite{BCV}.

We would like to diagonalize $H$ and find the
spectrum of this string. In the case of the maximal giant graviton ($p/N=1$)
 the Hamiltonian is integrable
 can be diagonalized using the Bethe ansatz. For $p \neq N$, we do not know at this moment
how to diagonalize (\ref{H}), and wether it is integrable or not. See however
\cite{Nepo} for a similar problem.

We have been able to find the ground state explicitly
\begin{eqnarray}
\label{ground} |\Psi_0\ket =(1-\alpha^2)^{L/2}\!\!\!\!\!\sum_{n_1,
\ldots, n_L = 0}^\infty \!\!\!\!\!\left(-\alpha\right)^{n_1 +
\cdots + n_L}|n_1,\ldots, n_L\ket\;,
\end{eqnarray}
 and it has energy $E = 0$.
 The expectation value of the number operator for the ground state
 is,
\begin{eqnarray}
\bra\Psi_0 | \hat{n} | \Psi_0\ket =   \frac{L N}{p} \left(1 -
\frac{p}{N} \right)\;,
\end{eqnarray}
which is generically of order $L$, unless $p<<N$.

Now we want to study the semiclassical limit for these open
strings (we abuse notation here because in the AdS/CFT
correspondence these represent strings attached to a giant
graviton).
 For that we need to take $L \sim \sqrt{N} \rightarrow
\infty$ (one also needs to take $\lambda\to \infty$ with
$\lambda/L^2$ fixed,  and we also keep $p/N$ fixed). We now need
to consider coherent states of the operators (\ref{ops}), along the lines of \cite{krusp}. Building
coherent states for this algebra is not trivial. The reason is
that the naive coherent states, $\hat{a} |z\ket = z|z\ket$ with
$|z\ket \sim \sum_n z^n |n\ket$, are not complete. Several
solutions to this problem has been proposed in the literature
\cite{minic, kar, klauder, baz}. For us this is not of much
concern because we are only interested in the classical action.
However, one can be more formal and define the overcomplete
coherent states along the lines of \cite{ baz} using the general
$q$-deformed algebra. Constructing the partition function then
goes in the familiar way \cite{zhang}. Taking the $L \rightarrow
\infty$ takes us to the classical limit. Therefore we can take $q
\rightarrow 0$ directly in the classical action.

In the $q \rightarrow 0$ limit, the coherent states of \cite{ baz}
reduce to the familiar form, $ |z\ket = \sqrt{1 - |z|^2} \sum_{n 0}^\infty z^n |n\ket\;, $ with $z \in \mathbb{C}$ and $|z| < 1$.
The coherent state for multiple sites can be written as $|z \ket
\equiv |z_1\ket \otimes \cdots \otimes |z_L\ket$. If we label the
coherent states as $z_l = r_l e^{i \phi_l}$ and take the continuum
limit $z_l(t) \rightarrow z(t,\sigma)$, we get the following
classical action,
\begin{eqnarray}
\label{action} S &=& \int dt \left(i \bra z|
\frac{\partial}{\partial t} |z \ket -
\bra z | H| z \ket \right) \nonumber \\
&=& -L \int dt \int_0^1 d\sigma \left[\frac{r^2 \dot{\phi}}{1 -
r^2}
+ \frac{\lambda}{L^2} (r'^2 + r^2 \phi'^2)\right] \nonumber \\
 &&- \left. \lambda \int dt\left[\alpha^2
\sin^2 \phi + \left(\alpha \cos \phi + r\right)^2 \right]
 \right|_{\sigma = 0} \nonumber \\
 &&- \left. \lambda \int dt\left[\alpha^2
\sin^2 \phi + \left(\alpha\cos \phi + r\right)^2 \right]
 \right|_{\sigma = 1}\;,
\end{eqnarray}
where the dot and primes are derivatives with respect to $t$ and
$\sigma$ respectively. Another approach is to consider our states
as a spin $j = 1/2$ representation of the $SL(2)$ algebra.
Building the coherent states goes as in \cite{bellucci}, and one
can recover the same classical action (\ref{action}) after some
field redefinitions.

 Ignoring the boundary terms (see below),
the equations of motion for the action (\ref{action}) are
\begin{eqnarray}
\label{eom1} \frac{r \dot{r}}{(1 - r^2)^2} + \frac{\lambda}{L^2}
\partial_\sigma(r^2 \phi'^2) &=&0 \;, \\
\frac{r \dot{\phi}}{(1 - r^2)^2} + \frac{\lambda}{L^2} (r \phi'^2
- r'') &=& 0\;.
\end{eqnarray}
The classical Hamiltonian for the coherent states is,
\begin{eqnarray}
\label{Hclass}
\bra H\ket &=& \frac{\lambda}{L} \int_0^1 d\sigma  (r'^2 + r^2 \phi'^2) \nonumber \\
&+& \left. \lambda \left[\alpha^2 \sin^2 \phi + \left(\alpha\cos
\phi + r\right)^2 \right]
 \right|_{\sigma = 0} \nonumber \\
&+& \left. \lambda \left[\alpha^2\sin^2 \phi + \left(\alpha\cos
\phi + r\right)^2 \right]
 \right|_{\sigma = 1}\;.
\end{eqnarray}
The average number of $Z$s in the open string is,
\begin{eqnarray}
\bra \hat{n} \ket = L \int_0^1 d\sigma \frac{r^2}{1 - r^2}\;,
\end{eqnarray}
and using Eq. (\ref{eom1}) we have,
\begin{eqnarray}
\label{ndot}
\partial_t \bra \hat{n} \ket = 2 \frac{\lambda}{L} \left(1 -
\frac{p}{N}\right) \left(\phi'|_{\sigma = 0} - \phi'|_{\sigma = 1}
\right)\;.
\end{eqnarray}
So in general $\bra \hat{n}\ket$ is not conserved and therefore
the string will oscillate in length. This is the way we measure
the length of the spin chain according to the XXX model. Note
however that we must ensure that $\bra \hat{n}\ket$ remains
bounded.

The boundary terms in the Hamiltonian (\ref{Hclass}) can give
rise to a large energy. From the point of view of the dual string theory, this
means that moving the ends of the open string costs a lot of
energy compared to the fluctuations of the bulk of the string.
Thus the lowest energy classical configurations will have  the
boundary terms in (\ref{Hclass}) set to zero. This gives rise to
the following Dirichlet boundary conditions:
\begin{eqnarray}
\label{bcr}
r|_{\sigma = 0,1} &=& \sqrt{1 - \frac{p}{N}}\;, \\
\label{bcsigma}
 \phi |_{\sigma
= 0,1} &=& \pi\;.
\end{eqnarray}

In fact, comparing with the results in the literature
\cite{mcgreeve} one can see that the boundary condition for $r$ is
just the radius of orbit of the giant graviton (in units of $R$).
On the other hand, the boundary condition on $\phi$ implies by
(\ref{ndot}) that, in general, the length of the string is not
fixed. The ground state will be  such that that $r'=\phi'=0$. This
is  a homogeneous string  and this corresponds to a massless
excitation of the brane.  Notice that this matches precisely the
form of the ground state we wrote, and this can be considered as a dual
derivation of the spectrum of massless excitations of the giant
graviton itself \cite{DJM}.

The spacetime interpretation of the boundary conditions
(\ref{bcr}) and (\ref{bcsigma}) can be made more clear  by
deriving the action (\ref{action}) directly as a limit of the
Polyakov action in a particular gauge. Since in our case the
number of $Y$s in the open string is constant ($ = L +1 \sim L$),
we should use a gauge that distributes the angular momentum in $Y$
homogeneously along the string, this gauge has also been considered in \cite{AFrol}. Moreover, the boundary condition
for $\phi$ suggests that we work in a frame where the giant
graviton is static.

To do this we follow \cite{kru} and write the Polyakov action in
momentum space,
\begin{eqnarray}
\label{paction} S_p = \sqrt{\lambda_{YM}} \int d\tau \int_0^\pi
\frac{d\sigma}{2 \pi} {\cal L} \;,
\end{eqnarray}
where,
\begin{eqnarray}
{\cal L} &=& p_\mu \partial_0 x^\mu + \frac{1}{2} A^{-1} \left[
G^{\mu \nu} p_\mu p_\nu + G_{\mu \nu} \partial_1 x^\mu
\partial_1 x^\nu \right] \nonumber \\
&&+ B A^{-1} p_\mu \partial_1 x^\mu\;.
\end{eqnarray}
Here we defined $\lambda_{YM} = g_{YM}^2 N = R^4/\alpha'^2$, and
$A$, $B$ play the role of Lagrange multipliers implementing the
constraints.

Now we write the metric of $\mathbb{R} \times S^5$ as,
\begin{eqnarray}
ds^2 = -dt^2  + |dX|^2 + |dY|^2 + |dZ|^2\;,
\end{eqnarray}
where $|X|^2 + |Y|^2 + |Z|^2 = 1$. The giant graviton is orbiting
in the $Z$ direction with $Z = \sqrt{1 - p/N} e^{i t}$ and wraps
the remaining $S^3$. We put our string at  $X = 0$ and define the
coordinates,
\begin{eqnarray}
Z = r e^{i (t  - \phi)}\;, \;\;\;\;\;\; Y = \pm \sqrt{1 - r^2}
e^{i \varphi}\;,
\end{eqnarray}
for which the giant graviton is static at $r = \sqrt{1 - p/N}$ and
$\phi$ constant. The metric becomes,
\begin{eqnarray}
\label{metric}
 ds^2 &=&  -(1 - r^2) dt^2 + 2 r^2 dt d\phi
+ \frac{1}{1 -
r^2} dr^2  \nonumber \\
&&+ r^2 d\phi^2 + (1 - r^2) d\varphi^2 \;.
\end{eqnarray}

The momentum in  $\varphi$ is conserved and is given by,
\begin{eqnarray}
 L  = \sqrt{\lambda_{YM}} \int_0^\pi \frac{d\sigma}{2 \pi}
p_\varphi \equiv \sqrt{\lambda_{YM}} {\cal J} \;.
\end{eqnarray}
We choose a gauge in which angular momentum $p_\varphi$ is
homogeneously distributed and $\tau$ coincides with the global
time in the metric,
\begin{eqnarray}
t = \tau \;, \;\;\;\;\;\; p_\varphi = 2 {\cal J} = {\rm const}.
\end{eqnarray}

We then implement the constraints that follow from varying $A$ and
$B$ in (\ref{paction}) directly in the action as done in \cite{kru}. Then, using
the equations of motion of  $p_r$ and $p_\phi$   the action is written  in terms
of the fields $r$ and $\phi$ and their derivatives. Finally, we
take the limit ${\cal J} \rightarrow \infty$ and assume that the
time derivatives are of the order $\partial_0 x^\mu \sim 1/{\cal
J}^2$.  To  order ${\cal O}(1/{\cal J}^2)$ ,
\begin{eqnarray}
S_p \approx - L \iint_0^\pi   \frac{dtd\sigma}{\pi} \left[
\frac{r^2 \dot{\phi}}{1-r^2} + \frac{1}{8 {\cal J}^2} (r'^2 + r^2
\phi'^2)   \right]\,.
\end{eqnarray}
Then, rescaling $\sigma \rightarrow \pi \sigma$ and using
$\lambda_{YM}/(8 \pi^2) = \lambda$, we get the action
(\ref{action}) (without the boundary terms).

Therefore, we see that the fields $r$ and $\phi$  of the coherent
states are the spacetime coordinates for an open string attached
to a giant graviton in a coordinate system for which the brane is
static. Furthermore, we see how the particular world-sheet gauge
is encoded in the CFT side: we have {\it chosen} to label our
states in such a way so that the $Y$s are distributed
homogeneously along the operator. This has very strong
implications in the AdS/CFT correspondence, because we are seeing
explicitly the reparametrization invariance of the string
worldsheet: the  gauge that makes the calculation more natural is
different than the one considered in other semiclassical setups \cite{kru,ft}.

S. E. V\'azquez would like to thank L. Balents for discussions. D.B. work was supported in part by a DOE OJI award, under grant DE-FG02-91ER40618 and NSF under grant No. PHY99-07949. D.H.C. work was supported in part by NSF under grant No. PHY99-07949 and by Fundaci\'on Antorchas. S.E.V. work was supported by an NSF graduate fellowship.

\end{document}